%Paper: hep-th/9511010
%From: roman@ecm.ub.es
%Date: Thu, 2 Nov 1995 10:18:42 +0100
%Date (revised): Thu, 2 Nov 1995 17:21:02 +0100

\magnification=1200
\parskip 10pt plus 1pt
\nopagenumbers

\def\r{{\rm I \hskip -2pt R}}
\def\p{{\rm I \hskip -2pt P}}

\overfullrule=0pt
\hfill UB-ECM-PF-95/21
%\vskip 5cm
%\bigskip\bigskip
\vskip 2cm
\centerline{\bf THE REGULATED FOUR PARAMETER}
\centerline{\bf ONE DIMENSIONAL POINT INTERACTION}
%\centerline{{\bf Coulomb-type bound states}\footnote{*}{This work is
%a reduced version of the preprint UB-ECM-PF-95/6,
%hep-ph/9505442 (unpublished) with the same title.}}
\vskip 1.5cm
\centerline{\it Jos\'e Mar\'{\i}a Rom\'an}
\vskip 0.2cm
\centerline{\it and}
\vskip 0.2cm
\centerline{\it Rolf Tarrach}
\vskip 0.5cm
\centerline{Departament d'Estructura i Constituents de la Mat\`eria}
\centerline{Facultat de F\'{\i}sica}
%\centerline{Institut de F\'\i sica d'Altes Energies}
%\smallskip
\centerline{Universitat de Barcelona}
\centerline{Diagonal 647, 08028 Barcelona, Spain }
\centerline{and}
\centerline{IFAE}
%\vskip 1cm
\vskip 3cm

\centerline{\bf Abstract}
\bigskip   \bigskip

{\it
The general four parameter point interaction in one dimensional quantum
mechanics is regulated. It allows the exact solution, but not the
perturbative one. We conjecture that this is due to the interaction not
being asymptotically free. We then propose a different breakup of
unperturbed
theory and interaction, which now is asymptotically free but leads to
the
same physics. The corresponding regulated potential can be solved both
exactly and perturbatively, in agreement with the conjecture.}

\vfill
\footnote{}{e-mail: roman@ecm.ub.es}
\footnote{}{e-mail: tarrach@ecm.ub.es}
\eject

\footline ={\hss\tenrm\folio\hss}
\pageno = 1

\beginsection {1. Introduction}

The easiest point interaction in one dimensional quantum mechanics was
introduced by Fermi more than three scores ago [1]. It corresponds to a
Dirac $\delta$, and its mathematical interpretation came almost thirty
years later [2]. It is now an almost standard example in elementary
quantum mechanics. In one dimension, however, and only in one dimension,
there exists a much more complex point interaction, which in its most
general form depends on four real parameters [3]. One of these
corresponds to the so-called $\delta^{\prime}$ potential, an
interaction
surrounded with confusion, controversy and issues of interpretation
[4-8]. There are of course no difficulties if the problem is solved in
terms of boundary conditions which make sure that the hamiltonian is
selfadjoint. Even its Brownian measure has been contructed [9]. A deeper
physical understanding, however, requires a regulated potential which,
when eventually the regulator is removed, leads to the same physics.
Surprisingly it does not seem to exist in the published literature, and
it even has been put forward that the problem does not allow regulation.
The nearest one has come is, on one hand, a regulated hamiltonian, which
however
is not selfadjoint, though it converges to one which is selfadjoint when
the short distance cutoff is removed [10]; and on the other, the proof
that a regulator exists for three parameters, which include the
so-called
$\delta^2$ [11]. This is, to a physicist, not a satisfactory state of
the art.

Here we give a complete solution to the problem. We find a regulated
potential which corresponds to a selfadjoint hamiltonian and reproduces
all the four-parameter physics. We then show that it however does not
allow to solve the problem perturbatively, i.e., it leads to a
perturbation theory which is not renormalizable. We argue that this is
because the interaction is not asymptotically free, as there is
scattering at infinite energy. We then reformulate the problem by
proposing a different partition of unperturbed hamiltonian and
interaction.
The new regulated potential allows now both an exact and perturbative
solution. This is consistent with our conjecture that asymptotic
freedom is necessary for perturbative renormalizability in quantum
mechanics, as the new interaction is asymptotically free.

We start with a section where the physics of the four parameter boundary
conditions, as well as all the limits and particular cases, are
reviewed. For the next two sections, which contain all the results,
references [12] might be helpful for those readers not used to the
language of quantum field theory as applied to quantum mechanics. Our
units are $\hbar=2m=1$.

\beginsection {2. The boundary conditions and its physics}

The most general point interaction in one dimension is described by a
free hamiltonian on the real line with the origin excluded where the
boundary values of the wavefunction and its derivative satisfy the
constraints [9]:

$$\pmatrix{-\psi^{\prime}_L\cr
            \psi^{\prime}_R\cr}=\pmatrix{\rho+\alpha  &
                       -\rho e^{i\theta}\cr
                       -\rho e^{-i\theta}  &
                        \rho+\beta\cr} \pmatrix{\psi_L\cr
                        \psi_R\cr} \eqno(2.1) $$

\noindent
which ensures selfadjointness of the hamiltonian. Here $\rho\ge0,~
\alpha,~ \beta$ and $0\le\theta<2\pi$ are real parameters and the
subindices
indicate whether the value of the wavefunction or its derivative, which
both have to be finite, correspond to the boundary of the negative
halfline (L) or the positive halfline (R). Eq. (2.1) can also be
written, for $\rho>0$, as

$$\pmatrix{\psi^{\prime}_R\cr
           \psi_R\cr}=
 e^{-i\theta}\pmatrix{1+\beta/\rho  &  \alpha+\beta+\alpha\beta/\rho \cr
                      1/\rho        &  1+\alpha/\rho                \cr}
             \pmatrix{\psi^{\prime}_L \cr
                      \psi_L       \cr} \eqno(2.2)$$

\noindent
which is more adequate for taking the $\rho \to \infty$ limit.

Notice that eqs. (2.1) and (2.2) are invariant under

$$\eqalign{
       \psi_R &\leftrightarrow \psi_L \cr
       \psi^{\prime}_R &\leftrightarrow -\psi^{\prime}_L  \cr}
                                                      \eqno(2.3)$$

\noindent
which corresponds to the parity operation, $x \leftrightarrow -x$,
together with

$$\eqalign{
         \alpha &\leftrightarrow \beta \cr
         \theta &\leftrightarrow -\theta \cr}
                                               \eqno(2.4)$$

\noindent
which will therefore correspond to changing the sign of the
antisymmetric interaction. This implies violation of parity invariance,
as long as $\alpha \neq \beta$ or $\theta \neq 0$.

For $\rho=0$ eq. (2.1) reduces to

$$\eqalign{
        -\psi^{\prime}_L &= \alpha\psi_L  \cr
         \psi^{\prime}_R &= \beta\psi_R   \cr}
                                              \eqno(2.5)$$

\noindent
which corresponds to a $L^2(\r^-)$ and $L^2(\r^+)$ problem respectively,
unrelated to each
other, with no probability flowing from one to the other. The parameter
$\theta$ becomes irrelevant. If furthermore $\alpha=\beta=0$ we have

$$\psi^{\prime}_L=\psi^{\prime}_R=0 \eqno (2.6)$$

\noindent
which are Neumann boundary conditions on both halflines.

For $\alpha=\beta,~ \theta=0$ eq. (2.1) becomes

$$\eqalign{
  \psi^{\prime}_R - \psi^{\prime}_L &= \alpha (\psi_R + \psi_L) \cr
  \psi^{\prime}_R + \psi^{\prime}_L &= (2\rho + \alpha)(\psi_R - \psi_L)
                                               \cr}    \eqno (2.7)$$

\noindent
which corresponds to a symmetric interaction. The first condition refers
to the even wavefunction, and the second to the odd. If furthermore
$\alpha=0$ then

$$\eqalign{
  \psi^{\prime}_R &= \psi^{\prime}_L \cr
  \psi^{\prime}_L &= \rho(\psi_R - \psi_L) \cr}
                                                 \eqno (2.8)$$

\noindent
which corresponds to the ill-called (it is symmetric!) $\delta^{\prime}$
interaction, which only acts on odd wavefunctions. If finally $\rho
\to \infty$ there is no interaction,

$$\eqalign{
  \psi^{\prime}_R &= \psi^{\prime}_L \cr
  \psi_R &= \psi_L \cr}
                                                  \eqno (2.9)$$

For $\rho \to \infty$ and $\theta=0$, eq. (2.2) reduces to

$$\eqalign{
  \psi_R &= \psi_L \cr
  \psi^{\prime}_R - \psi^{\prime}_L &= (\alpha + \beta) \psi_L \cr}
                                                    \eqno (2.10)$$

\noindent
which is the $\delta$ interaction. The parameter $\alpha - \beta$ is
irrelevant. It is also a symmetric interaction which only acts on even
wavefunctions.

For $\rho \to \infty$ and $\alpha + \beta = 0$ eq. (2.2) becomes

$$\eqalign{
  \psi^{\prime}_R &= e^{-i\theta} \psi^{\prime}_L \cr
  \psi_R &= e^{-i\theta} \psi_L \cr}
                                                  \eqno (2.11)$$

\noindent
which shows that $\theta$ is just a constant phase shift in crossing the
origin. The breaking of time reversal invariance due to the
noninvariance of the boundary conditions under complex conjugation is
made clear in these last equations, but is seen already in eq. (2.1).

For $\rho>0$, if two bound states exist their energies are given by

$$\sqrt{-E_0}= -\rho -{1\over2}(\alpha + \beta) \pm
               {1\over2}\sqrt{4\rho^2 + (\alpha - \beta)^2} > 0
                                                    \eqno (2.12)$$

\noindent
They do not depend on $\theta$. When only one bound state exists its
energy is given by the upper sign expression of eq. (2.12). For $\rho=0$
at most one bound estate exists, and its energy is

$$\sqrt{-E_0}=-\alpha > 0   \eqno (2.13)$$

\noindent
which requires $\alpha=\beta<0$, so that the eigenvalues in $\r^+$ and
$\r^-$ are the same.

For the scattering states, $E \equiv k^2 > 0$, and defining the
scattering amplitudes according to

$$\psi(x)= e^{ikx} + ie^{ik|x|}\bigl( f_+(k)\theta(x) +
                                      f_-(k)\theta(-x) \bigr)
                                                  \eqno (2.14) $$

\noindent
with $\theta(x)=0$ for $x<0$ and $\theta(x) + \theta(-x) = 1$, we
have

$$\eqalign{
   f_+(k) &= i + {2k\rho\over\Delta}e^{-i\theta} \cr
   f_-(k) &= {-i\over\Delta}
     \bigl(k^2 - ik(\alpha - \beta) + \rho(\alpha + \beta) + \alpha\beta
                                           \bigr)  \cr} \eqno (2.15)$$

\noindent
where

$$\Delta \equiv  k^2 + ik(2\rho + \alpha + \beta) -
                \rho(\alpha + \beta) - \alpha\beta  \eqno (2.16)$$

The high energy limit of the scattering amplitudes is

$$\eqalign{
   \lim_{k \to \infty}f_+(k) &= i \cr
   \lim_{k \to \infty}f_-(k) &= -i \cr} \eqno (2.17)$$

\noindent
so that there is scattering even at infinite energy, and the
wavefunction becomes

$$\lim_{k \to \infty} \psi(x) = 2\cos(kx) \theta(-x) \eqno(2.18)$$

\noindent
which corresponds to total reflection with Neumann boundary condition.

At low energies

$$\eqalign{
   \lim_{k \to 0}f_+(k) &= i \cr
   \lim_{k \to 0}f_-(k) &= i \cr} \eqno (2.19)$$

\noindent
and

$$\lim_{k \to 0} \psi(x) = 2i\sin(kx) \theta(-x) \eqno(2.20)$$

\noindent
which also corresponds to total reflection, but with Dirichlet boundary
condition.

For $\rho=0$ (2.15) becomes

$$\eqalign{
   f_+(k) &= i \cr
   f_-(k) &= -i{k-i\alpha \over k+i\alpha} \cr} \eqno(2.21)$$

\noindent
which again correspond to total reflection. The parameters $\beta$ and
$\theta$ are then of course irrelevant.

For $\alpha=\beta=\theta=0$ we obtain

$$f_+(k)= i{k \over k+2i\rho} = -f_-(k) \eqno (2.22)$$

\noindent
which is the $\delta^{\prime}$ scattering amplitude.

For $\rho \to \infty, ~ \theta=0$ one obtains the $\delta$ scattering
amplitude,

$$f_+(k)= -i{\alpha+\beta \over 2ik-(\alpha+\beta)} = f_-(k)
                                                      \eqno (2.23)$$

\noindent
Notice that now the scattering amplitudes vanish at high energies, i.e.,
the limits $\rho \to \infty$ and $k \to \infty$ do not commute.

The optical theorem reads

$$ 2Imf_+(k) = |f_+(k)|^2 + |f_-(k)|^2  \eqno (2.24)$$

\noindent
It will be convenient to introduce

$$\eqalign{
     f_s(k) &\equiv {1\over2}\bigl( f_+(k) + f_-(k) \bigr) \cr
     f_a(k) &\equiv {1\over2}\bigl( f_+(k) - f_-(k) \bigr) \cr}
                                                   \eqno (2.25)$$

\noindent
One can check that for a symmetric interaction, i.e., $\alpha=\beta, ~
\theta=0$, the optical theorem holds for the symmetric and antisymmetric
scattering amplitudes,

$$\eqalign{
  Imf_s(\alpha=\beta,\theta=0) &= |f_s(\alpha=\beta,\theta=0)|^2 \cr
  Imf_a(\alpha=\beta,\theta=0) &= |f_a(\alpha=\beta,\theta=0)|^2 \cr}
                                                     \eqno(2.26)$$

\noindent
which allows a straightforward introduction of phase shifts. This is
because a symmetric interaction in one dimension is equivalent to a
rotationally invariant interaction in more dimensions.

\beginsection {3. The strong regulated potential}

 From the work of Carreau [10] we are led to consider the following
potential:
$$V(x) =
        \theta(x+\epsilon)\theta(\epsilon-x)
        \left(W(\epsilon) + 2iR(\epsilon){d \over dx}\right)
        +\bigl(B(\epsilon)-iR(\epsilon)\bigr)\delta(x-\epsilon)
        +\bigl(A(\epsilon)+iR(\epsilon)\bigr)\delta(x+\epsilon)
                                                    \eqno(3.1)$$

\noindent
where $W(\epsilon), ~ R(\epsilon), ~ A(\epsilon)$ and $B(\epsilon)$
are real functions of the short distance regulator $\epsilon>0$, which
eventually is taken to be zero. The terms which contain $R(\epsilon)$
are imaginary and violate time reversal invariance. They will take care
of the $\theta$ parameter and depend on the momentum.
The hamiltonian

$$H =-{d^2 \over dx^2} + V(x) \eqno (3.2)$$

\noindent
is selfadjoint, and althought $V(x)$ contains Dirac deltas it is
sufficiently regulated to allow for its exact solution without further
regulation. One can regulate the Dirac deltas too, but this complicates
the analysis unnecessarily. One
can then show that one recovers (2.1) in the limit $\epsilon \to 0$ if

$$\eqalign{
  &W(\epsilon) = {\rho^2 \over 4}f^2(\rho \epsilon) \cr
  &R(\epsilon) = -{\theta \over 2\epsilon} \cr
  &A(\epsilon) = -{\rho \over 2}f(\rho \epsilon) + \rho + \alpha \cr
  &B(\epsilon) = -{\rho \over 2}f(\rho \epsilon) + \rho + \beta  \cr}
                                                   \eqno (3.3)$$

\noindent
with the function $f(x)$ given by the small $x$ condition

$$\eqalign{
  &\lim_{x \to 0} {\exp\bigl(xf(x)\bigr) \over f(x)} = 1 \cr
  &\lim_{x \to 0} xf(x) \to \infty  \cr}
                                                 \eqno (3.4)$$

\noindent
which implies, for small $x$,

$$f(x) \sim {-\ln x + \ln(-\ln x) \over x}  \eqno (3.5)$$

\noindent
Notice that terms subdominant with respect to the ones shown in eq.
(3.3) are irrelevant in the $\epsilon \to 0$ limit.

For finite $\rho$ one does not need the function $f(x)$ beyond eq.
(3.5),
but if one is interested in the limit $\rho \to \infty$ one would need
an $f(x)$ such that

$$\eqalign{
  &\lim_{x \to \infty} {\exp\bigl(xf(x)\bigr) \over f(x)} = 1  \cr
  &\lim_{x \to \infty} xf(x) \to \infty                        \cr}
                                                      \eqno (3.6)$$

\noindent
but no solution to this system exists, so that the regulated potential
$V(x)$ is only valid for finite $\rho$. The $\rho \to \infty$ limit has
to be taken after the continuum limit $\epsilon \to 0$ is taken. Notice
that also for the $\rho \to 0$ limit, $W(\epsilon), ~ A(\epsilon)$ and
$B(\epsilon)$ diverge, so that also the $\rho \to 0$ limit has to be
taken after the $\epsilon \to 0$ limit. Finally, if one is interested in
high energies, also the $k \to \infty$ limit has to be taken after the
regulator is removed, $\epsilon \to 0$, as $V(x)$ leads to no
scattering
at high energies, while eq. (2.1) shows scattering at high energies. The
reason why $V(x)$ does not scatter at high enough energy is that it
corresponds to wells and barriers of finite depth and height (recall
that
de Dirac $\delta$s can be regularized as well), which can be neglected
as compared to the kinetic energy as $k$ becomes larger and
larger. The presence of a contribution linear in $k$ does not spoil the
argument. In other words, at high energy the physics is determined by
the regulator, and we are only interested in regulator independent
physics.

The potential is symmetric for $\alpha = \beta, ~ \theta=0$ as then
$A(\epsilon)=B(\epsilon)$ and $R(\epsilon)=0$. It would be antisymmetric
for $W(\epsilon)=0$ and $A(\epsilon) = -B(\epsilon)$, but no values of
the parameters $\rho, ~ \alpha, ~ \beta$ and $\theta$ allow for an
antisymmetric potential. This is why a genuine $\delta^{\prime}$ does
not exist.

Consider now a perturbative expansion based on the Lippmann-Schwinger
equation for the hamiltonian of eq. (3.2) with $V(x)$ given by eq. (3.1)
but with $W(\epsilon), ~ R(\epsilon), ~ A(\epsilon)$ and $B(\epsilon)$
to be fixed in such a way that as $\epsilon \to 0$ one reproduces the
perturbative expansion of the scattering amplitudes $f_+(k)$ and $f_-(k)$
given in eqs. (2.15) and (2.16). As the free theory corresponds to
$\alpha=\beta=\theta=0, ~ \rho \to \infty$, one expands for small
$\alpha, ~ \beta$ and $\theta$ and large $\rho$. To first order one
obtains

$$\eqalign{
  f_+^{(1)}(k) &= -\theta - {\alpha + \beta \over 2k} + {k \over 2\rho}
                                                                   \cr
  f_-^{(1)}(k) &= -{\alpha + \beta \over 2k} - {k \over 2\rho} \cr}
                                                         \eqno(3.7)$$

\noindent
One can see that the second order term is imaginary.

 From the Lippmann-Schwinger equation

$$\psi(x)= e^{ikx} - \int dy G_k(x-y) V(y) \psi(y) \eqno (3.8)$$

\noindent
where $G_k(x)$ is the free outgoing propagator

$$G_k(x)={i\over 2k}e^{ik|x|}  \eqno (3.9)$$

\noindent
and eq. (2.14) one obtains the first Born approximation
$$\eqalign{
  f_+^{(1)}(k) &= 2\epsilon R_{(1)}(\epsilon) - {1\over 2k}
      \bigl(2\epsilon  W_{(1)}(\epsilon) + A_{(1)}(\epsilon) +
                                           B_{(1)}(\epsilon)\bigr) \cr
  f_-^{(1)}(k) &=
  -{1\over2k} \bigl( 2\epsilon W_{(1)}(\epsilon) + A_{(1)}(\epsilon) +
                                         B_{(1)}(\epsilon) \bigr)
   +i\epsilon \bigl( A_{(1)}(\epsilon) - B_{(1)}(\epsilon) \bigr) \cr
  &\phantom{=\ }
  +k\epsilon^2 \left( {2\over3}\epsilon W_{(1)}(\epsilon) +
          A_{(1)}(\epsilon) + B_{(1)}(\epsilon) \right) + O(k^2)  \cr}
                                                   \eqno (3.10)$$

\noindent
This implies, comparing to eq. (3.7), that in the $\epsilon \to 0$ limit
$$\eqalign{
  &R_{(1)}(\epsilon) = -{\theta \over 2\epsilon} \cr
  &2\epsilon W_{(1)}(\epsilon) + A_{(1)}(\epsilon) + B_{(1)}(\epsilon) =
                                                \alpha + \beta  \cr
  &\epsilon \bigl( A_{(1)}(\epsilon) -
                  B_{(1)}(\epsilon) \bigr) = 0   \cr}
                                                      \eqno (3.11)$$

\noindent
but the term linear in $k$ in eq. (3.7) cannot be obtained for $f_+(k)$
from eq. (3.10). We then will have to consider it a second order term
and thus,

$$\epsilon^2 \left( {2\over3}\epsilon W_{(1)}(\epsilon) +
                 A_{(1)}(\epsilon) + B_{(1)}(\epsilon) \right) = 0
                                                         \eqno(3.12)$$

\noindent
holds furthermore. The eq. (3.11) and (3.12) have solutions for
$W(\epsilon),
 ~ R(\epsilon), ~ A(\epsilon)$ and $B(\epsilon)$ to first order.

The second Born approximation gives, for the real part,

$$\eqalign{
  Re f_+^{(2)}(k) &= 2 \epsilon R_{(2)}(\epsilon) -
     {1 \over 2k}\bigl( 2\epsilon W_{(2)}(\epsilon) +
            A_{(2)}(\epsilon) + B_{(2)}(\epsilon) \bigr) \cr
     &\phantom{=\ } +
     {\epsilon \over k} \left( R^2_{(1)}(\epsilon) -
            {2 \over 3} \epsilon^2 W^2_{(1)}(\epsilon) -
\epsilon W_{(1)}(\epsilon) \bigl( A_{(1)}(\epsilon)
 + B_{(1)}(\epsilon) \bigr) -
            A_{(1)}(\epsilon) B_{(1)}(\epsilon) \right)   \cr
     &\phantom{=\ } +
     {4 \over 3} k \epsilon^3 \left(
            {2\over 5} \epsilon^2 W^2_{(1)}(\epsilon) +
 \epsilon W_{(1)}(\epsilon) \bigl( A_{(1)}(\epsilon)
 + B_{(1)}(\epsilon) \bigr) +
            2 A_{(1)}(\epsilon) B_{(1)}(\epsilon) \right)  \cr
     &\phantom{=\ } +
     O(k^2)                               \cr}    \eqno (3.13)$$
\noindent
and
$$\eqalign{
  Re f_-^{(2)}(k) &=
    -{1 \over 2k}\bigl( 2\epsilon W_{(2)}(\epsilon) +
            A_{(2)}(\epsilon) + B_{(2)}(\epsilon) \bigr) \cr
     &\phantom{=\ } -
     {\epsilon \over 2k} \biggl(
                      {4 \over 3}\epsilon^2 W^2_{(1)}(\epsilon) -
                      2R^2_{(1)}(\epsilon) +
                      4\epsilon W_{(1)}(\epsilon) B_{(1)}(\epsilon) \cr
             &\phantom{= -{\epsilon \over 2k} \biggl( {4 \over 3}
                                  \epsilon^2 W^2_{(2)}(\epsilon)\ \;} +
                      B^2_{(1)}(\epsilon) - A^2_{(1)}(\epsilon) +
                      2 A_{(1)}(\epsilon)B_{(1)}(\epsilon) \biggr) \cr
     &\phantom{=\ } +
     k\epsilon^2 \left( {2\over3}\epsilon W_{(2)}(\epsilon) +
                      A_{(2)}(\epsilon) + B_{(2)}(\epsilon) \right) \cr
     &\phantom{=\ } +
     {1 \over 3}k \epsilon^3 \biggl(
                      {4 \over 5} \epsilon^2 W^2_{(1)}(\epsilon) -
                      2R^2_{(1)}(\epsilon) +
                      4\epsilon W_{(1)}(\epsilon) B_{(1)}(\epsilon) \cr
             &\phantom{= +{1 \over 3}\epsilon^3k \biggl( {4 \over 5}
                                \epsilon^2 W^2_{(2)}(\epsilon)\ \;} +
                      B^2_{(1)}(\epsilon) - A^2_{(1)}(\epsilon) +
                      2 A_{(1)}(\epsilon)B_{(1)}(\epsilon) \biggr) \cr
     &\phantom{=\ } +
     O(k^2)       \cr}                             \eqno (3.14)$$

\noindent
where the possible second order counterterms of the first Born
approximation
have been included, as corresponds to a renormalizable pertubation
theory. As the second order contributions to the scattering
amplitudes are purely imaginary, eqs. (3.13) and (3.14) are bound to
just reproduce the terms linear in $k$ of eq. (3.7). This implies, in
the $\epsilon \to 0$ limit

$$\epsilon R_{(2)}(\epsilon) = 0$$

$$\eqalign{
  &{1 \over 2}\bigl( 2\epsilon W_{(2)}(\epsilon) +
                    A_{(2)}(\epsilon) + B_{(2)}(\epsilon) \bigr) = \cr
        &\phantom{ {1 \over 2}\bigl( 2\epsilon W_{(2)}(\epsilon) + }
            \epsilon \biggl( R^2_{(1)}(\epsilon) -
            {2 \over 3} \epsilon^2 W^2_{(1)}(\epsilon) -
            \epsilon W_{(1)}(\epsilon) \bigl( A_{(1)}(\epsilon) +
                                              B_{(1)}(\epsilon) \bigr) -
            A_{(1)}(\epsilon) B_{(1)}(\epsilon) \biggr)           \cr}$$

$${4 \over 3} \epsilon^3 \left(
            {2\over 5} \epsilon^2 W^2_{(1)}(\epsilon) +
            \epsilon W_{(1)}(\epsilon) \bigl( A_{(1)}(\epsilon) +
                                              B_{(1)}(\epsilon) \bigr) +
            2 A_{(1)}(\epsilon) B_{(1)}(\epsilon) \right) =
        {1 \over 2\rho}
                                                        \eqno (3.15)$$

\noindent
from eq. (3.13) and
$$\eqalign{
  &\bigl( 2\epsilon W_{(2)}(\epsilon) +
         A_{(2)}(\epsilon) + B_{(2)}(\epsilon) \bigr) = \cr
  &-\epsilon  \left( {4 \over 3}\epsilon^2 W^2_{(1)}(\epsilon) -
                    2R^2_{(1)}(\epsilon) +
                    4\epsilon W_{(1)}(\epsilon) B_{(1)}(\epsilon) +
                    B^2_{(1)}(\epsilon) - A^2_{(1)}(\epsilon) +
                    2 A_{(1)}(\epsilon)B_{(1)}(\epsilon) \right) \cr}$$
$$\eqalignno{
  &-{1 \over 2\rho} =
  \epsilon^2 \left( {2\over3}\epsilon W_{(2)}(\epsilon) +
                    A_{(2)}(\epsilon) + B_{(2)}(\epsilon) \right) +
                                                         &(3.16) \cr
  &{1 \over 3}\epsilon^3 \left(
                    {4 \over 5}\epsilon^2 W^2_{(1)}(\epsilon) -
                    2R^2_{(1)}(\epsilon) +
                    4\epsilon W_{(1)}(\epsilon) B_{(1)}(\epsilon) +
                    B^2_{(1)}(\epsilon) - A^2_{(1)}(\epsilon) +
                    2 A_{(1)}(\epsilon)B_{(1)}(\epsilon) \right) \cr}$$

\noindent
from eq. (3.14). One can convince oneself that the system given by eqs.
(3.11), (3.12), (3.15) and (3.16) has no solution. Perturbation theory
starting from eq. (3.2) does not work; it is not renormalizable.

One can understand why. Perturbation theory in $1 \over \rho$ is based
on the fact that the theory is free for large $\rho$ (and
$\alpha=\beta=\theta=0)$. Recall, however that the exact regulated
potential whose form we are using in this perturbative approach gives
the correct large $\rho$ behaviour if first the regulator is removed
taking $\epsilon \to 0$. But perturbation theory is taking these limits
in reversed order, it assumes large $\rho$ for fixed $\epsilon$. The
non-renormalizability just reflects the non-commutativity of the $\rho
\to \infty, ~ \epsilon \to 0$ limits.

A telltale signal that perturbation theory is not renormalizable is
given by the fact that the theory is not asymptotically free, in other
words, that it interacts even at infinite energy. This is saying that
the potential diverges very strongly (stronger than a Dirac delta) as
$\epsilon \to 0$. It is too strong to allow iterations as are done in
perturbation theory.

This leads to the question of whether we are facing a genuinely
non-perturbative problem, like tunneling, or whether a pertubative
approach starting from a different decomposition of unperturbed
hamiltonian and perturbation would allow a meaningfull perturbation
theory. Let us show that the second scenario holds.

\beginsection {4. The weak regulated potential}

We have conjectured that asymptotic freedom is essential for
perturbative
renormalizability in quantum mechanics. This leads us to trying to shift
the high energy scattering into the free part of the hamiltonian, so
that
the new interaction is asymptoticaly free. If our conjecture is correct
this should then allow a regularization which works both exactly and
perturbatively. This is indeed the case. Let us show how.

If the high energy behaviour of the scattering amplitudes, eq. (2.17),
is subtracted, eq. (2.14) reads

$$\psi(x) = 2\cos(kx) \theta(-x) + ie^{ik|x|}
        \bigl( f_s(k) + (f_a(k) - i) \epsilon(x) \bigr)
                                                        \eqno (4.1)$$

\noindent
where $\epsilon(x) \equiv \theta(x) - \theta(-x)$ and eq. (2.25) has
been used. Eq. (4.1) is not a good starting point for perturbation
theory, because the unperturbed wave vanishes on $\r^+$. Let us
therefore
transform eq. (4.1) under parity and under the substitution of eq.
(2.4),

$$\tilde \psi(-x) = 2\cos(kx) \theta(x) + ie^{ik|x|}
        \bigl( \tilde f_s(k) - (\tilde f_a(k) - i) \epsilon(x) \bigr)
                                                        \eqno (4.2)$$

\noindent
where $\tilde f(\rho,\alpha,\beta,\theta) =
f(\rho,\beta,\alpha,-\theta)$.
By summing and subtracting eq. (4.1) and eq. (4.2) one obtains
$$\eqalignno{
  \psi^{(+)}(x) &= \cos(kx) + ie^{ik|x|}
        \bigl( f^{(+)}_s(k) + f^{(+)}_a(k) \epsilon(x) \bigr)
                                                        &(4.3) \cr
&\cr
  \psi^{(-)}(x) &= -\cos(kx) \epsilon(x) + ie^{ik|x|}
        \bigl( f^{(-)}_s(k) + f^{(-)}_a(k) \epsilon(x) \bigr)
                                                        &(4.4) \cr}$$

\noindent
where
$$\eqalignno{
  f^{(+)}_s(k) &\equiv
               {1 \over 2} \bigl( f_s(k) + \tilde f_s(k) \bigr) =
    {1 \over \Delta} \left( k\rho (\cos \theta - 1) -{k \over 2}
    (\alpha + \beta) - i\rho(\alpha + \beta) - i\alpha\beta \right) \cr
  f^{(+)}_a(k) &\equiv
               {1 \over 2} \bigl( f_a(k) - \tilde f_a(k) \bigr) =
{k \over \Delta} \left( -i\rho \sin \theta + {1 \over 2}
                                        (\alpha - \beta) \right) \cr
& &(4.5) \cr
  f^{(-)}_s(k) &\equiv
               {1 \over 2} \bigl( f_s(k) - \tilde f_s(k) \bigr) =
{k \over \Delta} \left( -i\rho \sin \theta - {1 \over 2}
                                        (\alpha - \beta) \right) \cr
  f^{(-)}_a(k) &\equiv
              {1 \over 2} \bigl( f_a(k) + \tilde f_a(k) \bigr) -i =
    {1 \over \Delta} \left( k\rho (\cos \theta + 1) +{k \over 2}
    (\alpha + \beta) + i\rho(\alpha + \beta) + i\alpha\beta \right)
\cr}$$
% \eqno (4.5)$$

\eject
\noindent
Notice that the new scattering amplitudes all vanish at high energies.
The new unperturbed solutions are characterized by having vanishing
derivatives at the origin, as corresponds to
$\rho=\alpha=\beta=\theta=0$; recall eq. (2.6).

Looking at the wavefuctions as given by eqs. (4.3) and (4.4) for each
parameter alone we notice that they correspond to $\delta$-interactions.
So we expect that Dirac deltas appear in some sense in the potential.

We will therefore regularize the interaction with Dirac deltas away from
the origin. Having in mind that Dirac deltas can be characterized in
terms of a boundary condition, as shown in eq. (2.10), which we write as

$$\Delta \psi^\prime(0) = \lambda \psi(0) \quad  \longleftrightarrow
                    \quad  \lambda \delta(x) \psi(x)
                                                    \eqno (4.6)$$

\noindent
we now rewrite eq. (2.1) in a form which resembles eq. (4.6). In order
to do so, let us split it into conditions on two boundaries, writing
$\psi_L \to \psi(-\epsilon), ~ \psi_R \to \psi(\epsilon), ~
\psi^\prime_L \to -\Delta \psi(-\epsilon)$ and $\psi^\prime_R \to
\Delta \psi(\epsilon)$, where we have taken $\psi^\prime(0)=0$. Then eq.
(2.1) is substituted by

$$\eqalign{
  &\Delta\psi^\prime(-\epsilon) = (\rho + \alpha)\psi(-\epsilon)
                                  -\rho e^{i\theta}\psi(\epsilon) \cr
  &\Delta\psi^\prime(\epsilon) =  -\rho e^{-i\theta}\psi(-\epsilon)
                                  +(\rho + \beta)\psi(\epsilon)   \cr}
                                                         \eqno (4.7)$$

\noindent
In the limit $\epsilon \to 0$ eq. (4.7) coincides with eq. (2.1)

Comparing eqs. (4.7) and (4.6) leads immediately to the Schr\"odinger
equation

$$\eqalign{
  -{d^2 \over dx^2}\psi(x)
  &+ \bigl( (\rho + \beta)\delta(x - \epsilon) +
           (\rho + \alpha)\delta(x + \epsilon) \bigr) \psi(x) \cr
  &+ \bigl( -\rho e^{-i\theta}\delta(x - \epsilon) -
             \rho e^{i\theta}\delta(x + \epsilon) \bigr) \psi(-x) =
  E\psi(x)                                                      \cr}
                                                     \eqno (4.8)$$

\noindent
subject to the boundary condition

$$\psi^\prime(0) = 0  \eqno (4.9)$$

Notice that eq. (4.8) is non-local. This non-locality is however
avoided by just re- writing (4.8) in terms of the parity
eigenfunctions,
i.e., $\psi_s(x) \equiv {1 \over 2} \bigl( \psi(x) + \psi(-x) \bigr)$
and $\psi_a(x) \equiv {1 \over 2} \bigl( \psi(x) - \psi(-x) \bigr)$.
Then eq. (4.8) becomes

$$\left[ -{d^2 \over dx^2} +
         \pmatrix{ V_{ss}(x)  &  V_{sa}(x)  \cr
                   V_{as}(x)  &  V_{aa}(x)  \cr}
                                                  \right]
  \pmatrix{ \psi_s(x)  \cr
            \psi_a(x)  \cr} =
  E \pmatrix{ \psi_s(x)  \cr
                \psi_a(x)  \cr}
                                                      \eqno (4.10)$$

\noindent
where the potentials are given by

$$\eqalign{
V_{ss}(x) &=
{1\over2} \bigl( \alpha + \beta + 2\rho(1-\cos\theta) \bigr)
\bigl( \delta(x - \epsilon) + \delta(x + \epsilon) \bigr)    \cr
V_{aa}(x) &=
{1\over2} \bigl( \alpha + \beta + 2\rho(1+\cos\theta) \bigr)
\bigl( \delta(x - \epsilon) + \delta(x + \epsilon) \bigr)    \cr
V_{sa}(x) &=
-{1\over2} \bigl( \alpha - \beta + 2i\rho\sin\theta \bigr)
 \bigl( \delta(x - \epsilon) - \delta(x + \epsilon) \bigr)
                                            = V_{as}^{\ast}(x) \cr}
                                                       \eqno (4.11)$$

\noindent
together with

$$\psi^{\prime}_s(0) = \psi^{\prime}_a(0) = 0 \eqno (4.12)$$

\noindent
The potential matrix has been written in such a way that $V_{ss}(x)$
and $V_{aa}(x)$ are symmetric and $V_{sa}(x)=V_{as}^{\ast}(x)$ are
antisymmetric.
The solutions of eq. (4.10) reproduces eqs. (4.3) and (4.4) in the limit
$\epsilon \to 0$.

The interaction regulated in this way is actually a problem in the
Hilbert space $L^2(\r^+ \oplus \r^-)$ instead of $L^2(\r)$ with
potentials
acting on each halfline. The complex potentials are responsible
for the probability flow between the halflines. This topological feature
of the contact interaction is of course unique to one dimension, it
does not happen in higher dimensions.

One can rewrite eq. (4.8) with the help of the parity operator, $\p$,
and eq. (4.11) as
$$\eqalign{
  -{d^2 \over dx^2}\psi(x)
  &+ {1 \over 2} \bigl[ \bigl( V_{ss}(x) + V_{aa}(x) \bigr) +
             \bigl( V_{as}(x) + V_{sa}(x) \bigr) \bigr] \psi(x)  \cr
  &+ {1 \over 2} \bigl[ \bigl( V_{ss}(x) - V_{aa}(x) \bigr) +
             \bigl( V_{as}(x) - V_{sa}(x) \bigr) \bigr]\p \psi(x) =
  E\psi(x)                                                       \cr}
                                                         \eqno (4.13)$$

It might seem surprising that now the parity operator appears in the
regulated potential, whereas the momentum operator appeared in eq.
(3.1).
This is because the momentum operator is not regularized when acting
together with Dirac deltas and if the Dirac deltas are further
regularized, then it would act at the origin, where the wavefunction is
allowed to be discontinuous. The parity operator avoids these problems
and still allows violation of time reversal invariance, as can be seen
from the term $\bigl( V_{as}(x) - V_{sa}(x) \bigr)\p$, which is
imaginary.

To perform pertubation theory starting from the Lippmann-Schwinger
equation we need the propagator corresponding to the unperturbed
solutions of eqs. (4.3) and (4.4)

$$G_k(x,y)={i\over2k} \left(e^{ik|x+y|}+e^{ik|x-y|}\right)
         \bigl( \theta(x)\theta(y) + \theta(-x)\theta(-y) \bigr)
                                                        \eqno(4.14)$$

\noindent
which represents outgoing propagation which does not cross the origin
and which satisfies the adequate boundary conditions at the origin,
i.e., vanishing derivatives.

Using eqs. (4.11) and (4.14) one writes the Lippmann-Schwinger equation
for $\psi^{(+)}(x)$
$$\pmatrix {\psi^{(+)}_s(x)  \cr
            \psi^{(+)}_a(x)  \cr} =
   \pmatrix{ \cos(kx)   \cr
                0       \cr}
  -\int dy G_k(x,y)
        \pmatrix{ V_{ss}(y)   &   V_{sa}(y)   \cr
                  V_{as}(y)   &   V_{aa}(y)   \cr}
        \pmatrix {\psi^{(+)}_s(y)  \cr
                  \psi^{(+)}_a(y)  \cr}
                                                       \eqno (4.15)$$

\noindent
which allows to perform the perturbative expansion in a straightforward
way. This expansion can immediately be summed, and it reproduces the
exact result (4.3) in the limit $\epsilon \to 0$. The same can be done
for $\psi^{(-)}(x)$
with the same potential and we obtain the result of eq. (4.4).

Notice that for the $\delta^{\prime}$, $\alpha = \beta = \theta = 0$,
only $V_{aa}(x)$ remains. It is symmetric but only acts on the odd
wavefunctions.

The four parameter contact interaction is finally very elementary, as
seen in eq. (4.11), once the unperturbed hamiltonian and its acompanying
boundary conditions are properly chosen.

\beginsection {5. Conclusion}

The most general point interaction in one dimensional quantum mechanics
depends on four real parameters which determine the boundary conditions
at the site of the interaction. We here present, for the first time, a
regulated potential which leads to a selfadjoint hamiltonian and which
reproduces the same physics when the regulator is removed. Perturbation
theory built upon this regulated potential is however not
renormalizable. The regulator cannot be removed. We conjecture that
this is because the interaction is not asymptotically free, i.e.,
because
there exists scattering at infinite energy. This reflects too strong a
potential, which thus cannot be iterated perturbatively.

We then present a different breakup of unperturbed hamiltonian and
interaction,
for which the interaction is asymptotically free. We give its regulated
form, which allows both the exact and the perturbative solution. This
provides support to our conjecture.

\beginsection {Acknowledgments}

We thank Pedro Pascual and Josep Taron for reading the manuscript and
for their constructive comments. R. T. thanks Carlos Garc\'{\i}a-Canal,
Huner Fanchiotti and Luis Epele for hospitality and discussions at the
Universidad de La Plata, Argentina and Aneesh
Manohar for hospitality at the Dept. of Physics at UCSD, where this work
was partially performed. The Ph.D. thesis of Alejandro Rivero
has been a very useful source of references. Financial support under
CICYT contract AEN95-0590, DGICYT contract PR95-015 and
CIRIT contract GRQ93-1047 is acknowledged. J. M. R. is supported by a
Basque Country Goverment F. P. I. grant. He thanks Joan Soto for a
useful conversation, Josep Herrero for a good instruction
in Mathematica, and Assumpta Parre\~no for interesting comments.

\beginsection {References}

\item{1.} E. Fermi, Nuovo Cimento {\bf 11}, 157 (1934).
\item{2.} F. A. Berezin and L. D. Faddeev, Sov. Math. Dokl. {\bf 2}, 372
(1961).
\item{3.} S. Albeverio, F. Gesztesy, R. Hoegh-Krohn and H. Holden,
"Solvable models in quantum mechanics", Springer-Verlag, Berlin, 1988.
\item{4.} P. \v Seba, Rep. Math. Physics {\bf 24}, 111 (1986).
\item{5.} B.-H. Zhao, Jour. Physics {\bf A25}, L617 (1992).
\item{6.} D. J. Griffiths, Jour. Physics {\bf A26}, 2265 (1993).
\item{7.} S. Albeverio, F. Gesztesy and H. Holden, Jour. Physics
{\bf A26}, 3903 (1993).
\item{8.} P. B. Kurasov, A. Scrinzi and N. Elander, Phys. Rev.
{\bf A49}, 5005 (1994).
\item{9.} M. Carreau, E. Fahri and S. Gutmann, Phys. Rev. {\bf D42},
1194 (1990).
\item{10.} M. Carreau, Jour. Physics {\bf A26}, 427 (1993).
\item{11.} P. R. Chernoff and R. J. Hughes, Jour. Funct. Anal.
{\bf 111}, 97 (1993).
\item{12.} P. Gosdzinsky and R. Tarrach, Am. Jour. Physics {\bf 59}, 70
(1991).
\item{}    L. R. Mead and J. Godines, Am. Jour. Physics {\bf 59}, 935
(1991).
\item{}    R. Jackiw, in B\'eg Memorial Volume, ed. A. Ali and P.
Hoodbhoy, World Scientific, Singapore, 1991.

\end